\RequirePackage{fix-cm}
\documentclass[twocolumn]{svjour3}          
\smartqed  
\usepackage{graphicx}
\usepackage{epstopdf}
\usepackage{graphicx,color}

\begin{document}
\title{Multiple Andreev reflections spectroscopy of two-gap 1111- and 11 Fe-based superconductors}

\author{Ya.G.~Ponomarev \and
S.A.~Kuzmichev \and
T.E.~Kuzmicheva \and
M.G.~Mikheev \and
M.V.~Sudakova \and
S.N.~Tchesnokov \and
O.S.~Volkova \and
A.N.~Vasiliev \and
V.M.~Pudalov \and
A.V.~Sadakov \and
A.S.~Usol'tsev \and
Th.~Wolf \and
E.P.~Khlybov \and
 L.F.~Kulikova
}

\institute{Ya.G.~Ponomarev \and
S.A.~Kuzmichev \and
T.E.~Kuzmicheva \and
M.G.~Mikheev  \and
M.V.~Sudakova \and
S.N.~Tchesnokov \and
O.S.~Volkova \and
A.N.~Vasiliev \at
Lomonosov Moscow State University, 119991 Moscow, Russia
\and
T.E.~Kuzmicheva \and
V.M.~Pudalov \and
A.V.~Sadakov \and
A.S.~Usol'tsev \at
Lebedev Physical Institute RAS, 119991 Moscow, Russia
\and
Th.~Wolf \at
Karlsruher Institut f\"{u}r Technologie, Institut f\"{u}r Festk\"{o}rperphysik, D-76021 Karlsruhe, Germany
\and
E.P.~Khlybov  \and
L.F.~Kulikova \at
Institute for High Pressure Physics RAS, 142190 Troitsk, Russia
}

\date{Received: date / Accepted: date}

\maketitle

\begin{abstract}
Using the ``break-junction'' technique we prepared and studied superconductor - constriction - superconductor nanocontacts in polycrystalline samples of Fe-based superconductors CeO$_{0.88}$F$_{0.12}$FeAs (Ce-1111; $T_C^{\rm bulk} = 41 \pm 1$\,K), LaO$_{0.9}$F$_{0.1}$FeAs (La-1111; $T_C^{\rm bulk} = 28 \pm 1$\,K), and FeSe ($T_C^{\rm bulk} = 12 \pm 1$\,K). We detected two subharmonic gap structures related with multiple Andreev reflections, indicating the presence of two superconducting gaps with the BCS-ratios $2\Delta_L/k_BT_C = 4.2 \div 5.9$ and $2\Delta_S/k_BT_C\sim 1 \ll 3.52$, respectively. Temperature dependences of the two gaps $\Delta_{L,S}(T)$ in FeSe indicate a $k$-space proximity effect between two superconducting condensates. For the studied iron-based superconductors we found a linear relation between the gap $\Delta_L$ and magnetic resonance energy, $E_{\rm res} \approx 2\Delta_L$.
\keywords{Fe-based superconductors \and two-gap superconductivity \and multiple Andreev reflections \and subharmonic gap structure \and ``break-junction''}
\PACS{74.70.Xa \and 74.25.-q \and 74.45.+c}
\end{abstract}

\section{Introduction}
\label{intro}

Andreev spectroscopy \cite{Andreev} is a powerful instrument to measure superconducting gap in a wide temperature range, up to $T_C$ \cite{Kummel,Poenicke,Blonder}. A number of such measurements have been performed earlier with oxypnictides of the RFeAsO$_{1-x}$F$_x$ family and with FeSe \cite{Seidel,LOFA,Gd,FeSe}. Here we present systematic studies of the current-voltage characteristics (CVCs) and dynamic conductance $dI(V)/dV$ for superconductor - constriction - superconductor (ScS) contacts in Ce-1111, La-1111 and FeSe. Using the intrinsic multiple Andreev reflections effect (IMARE) spectroscopy, we  measured the two superconducting gap values in all three Fe-based materials and determined temperature dependences of the two gaps for FeSe.

The compounds under study belong to the class of iron-based superconductors discovered in 2008 \cite{Kamihara}. These materials are characterized by a layered crystal structure; their electron energy spectrum in the normal state contains electron and hole quasi-two-dimensional Fermi surface sheets, where two superconducting condensates are supposed to be formed at $T < T_C$ \cite{Seidel}.

\section{Experimental details}
\label{exp}

To measure the superconducting gaps we used two methods: (i) Andreev spectroscopy \cite{Andreev} of single ScS nanocontacts \cite{Kummel}, and (ii) IMARE spectroscopy of ScS-contact stacks. The nano-sized contacts required for multiple Andreev reflections spectroscopy, have been made using the ``break-junction'' technique \cite{Moreland}. In this technique, breaking a bulk sample in the cryogenic environment creates a superconductor - constriction - superconductor (ScS) junctions. Bias current flowing through the sub-mcm size constriction exceeds the superconducting critical current value and causes the contact area transition to the normal state; as a result, the studied ScS-contacts may be considered as conventional SnS-junctions.

The main features of the $I(V)$ curves for our ScS-contacts comprise a pronounced excess current at low bias voltages and a subharmonic gap structure (SGS) in the $dI/dV$ curve. The latter shows sharp dips at a set of bias voltages $V_n$. For interpreting these dips we use theoretical model by K\"{u}mmel \emph{et al.} \cite{Kummel}, applicable for conductance spectra of the symmetric ScS-contacts:

\begin{equation}
V_n = \frac{2\Delta}{en},~~~~~ n=1,2\dots
\end{equation}

As the subharmonic number $n$ increases, the dip amplitude decays. By plotting the $V_n(1/n)$ dependence (which must pass through the $(0;0)$ point) it is easy to determine the gap value accurately. In the case of a two-band superconductor, two distinct SGS should be observed.

Due to the local character of the Andreev spectroscopy of ScS break junctions, studies of the SGS for individual Sharvin type \cite{Sharvin} nanocontacts allow to gain information even in  case of inhomogeneous samples. In order to observe SGS, the size $a$ of the Andreev contact should be significantly smaller than the quasiparticles mean free path $l$ (the ballistic regime) \cite{Kummel,Poenicke,Blonder}.

Because of the layered structure of Fe-based superconductors, exfoliation of the sample generates nanosteps and terraces in the $c$-direction and thus may form not only single ScS-junctions but also arrays of the S-c-S-c-...-S- type junctions. The array represents a stack  of several  consequently connected ballistic ScS-junctions causing an intrinsic multiple Andreev reflections effect. The latter is similar to the intrinsic Josephson effect in SIS-array \cite{Nakamura}. Using stacks of contacts, one can exclude surface distortion of superconductivity and observe sharp peculiarities corresponding to {\it the true bulk gaps}. Bias voltages for these singularities should scale with the number of contacts $N$ in the stack.

For temperatures up to $T_C$, the gap $\Delta$ may be obtained directly by substituting to Eq.(1) the bias voltages corresponding to the dips \cite{Kummel}. Our data for three different materials  are shown in Figs.~1-5; they are typical for clean classical SnS-contacts \cite{Kummel}. As will be shown below, the data manifest two distinct sequences of dips.

The $I(V)$ and dI(V)/dV-characteristics were measured by a computer controlled set-up using a 16 bit National Instrument board. The dynamic conductance spectra $dI(V)/dV$ were measured by a standard modulation technique \cite{Rakhmanina}.

\section{Experimental results}
\label{results}
\subsection{CeO$_{1-x}$F$_x$FeAs}
\label{Ce}

In this section we present Andreev spectroscopy data for Ce-1111 break junctions.  The results reveal the existence of two superconducting energy gaps and enable evaluating their magnitude at $T = 4.2$\,K.
To the best of our knowledge, the gap values were not measured for this material earlier. The polycrystalline CeO$_{0.88}$F$_{0.12}$FeAs samples with $T_C^{\rm bulk} \approx 41$\ were synthesized as described in \cite {Khlybov}.

\begin{figure}
\begin{center}
\includegraphics[width=0.46\textwidth]{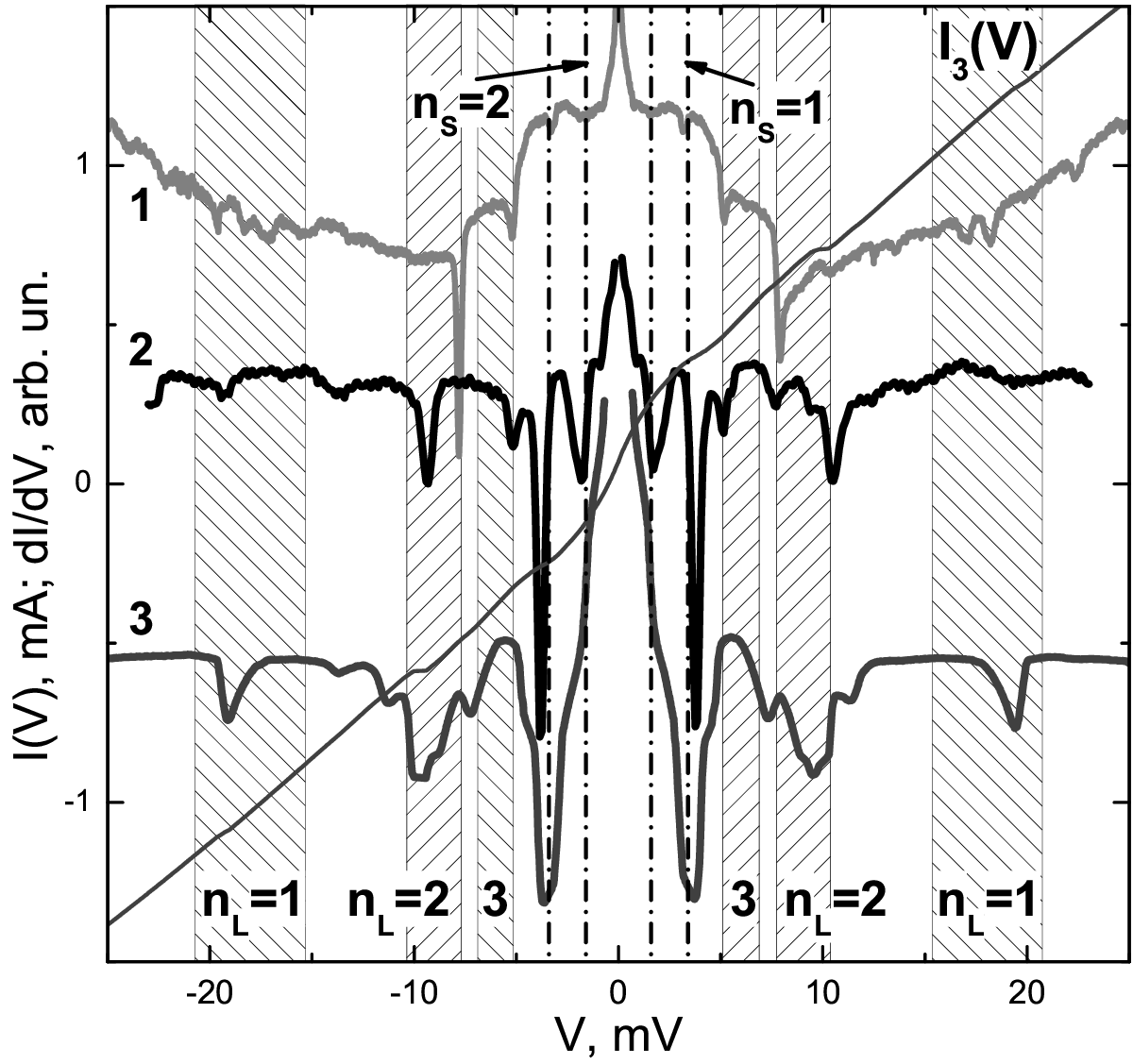}
\caption{ Ce-1111.
$dI/dV$-characteristics of three representative  ScS-contacts at $T=4.2$K: sample Ce1, contacts $\sharp d1$ (1) and $\sharp d2$ (2), and sample Ce2, contact $\sharp d6$ (3). Thin line I$_3$(V) shows CVC for the latter contact. The data reveal two sets of SGS corresponding to the gaps $\Delta_L \approx 9$\,meV (marked with $n_L$ labels), and $\Delta_S \approx 1.6$\,meV (dash-dotted vertical lines with $n_S$ labels). Dashed areas cover a 15\% uncertainty for the large gap value. The curves are shifted vertically, for clarity
}
\label{Fig1}
\end{center}
\end{figure}

\begin{figure}
\begin{center}
\includegraphics[width=0.42\textwidth]{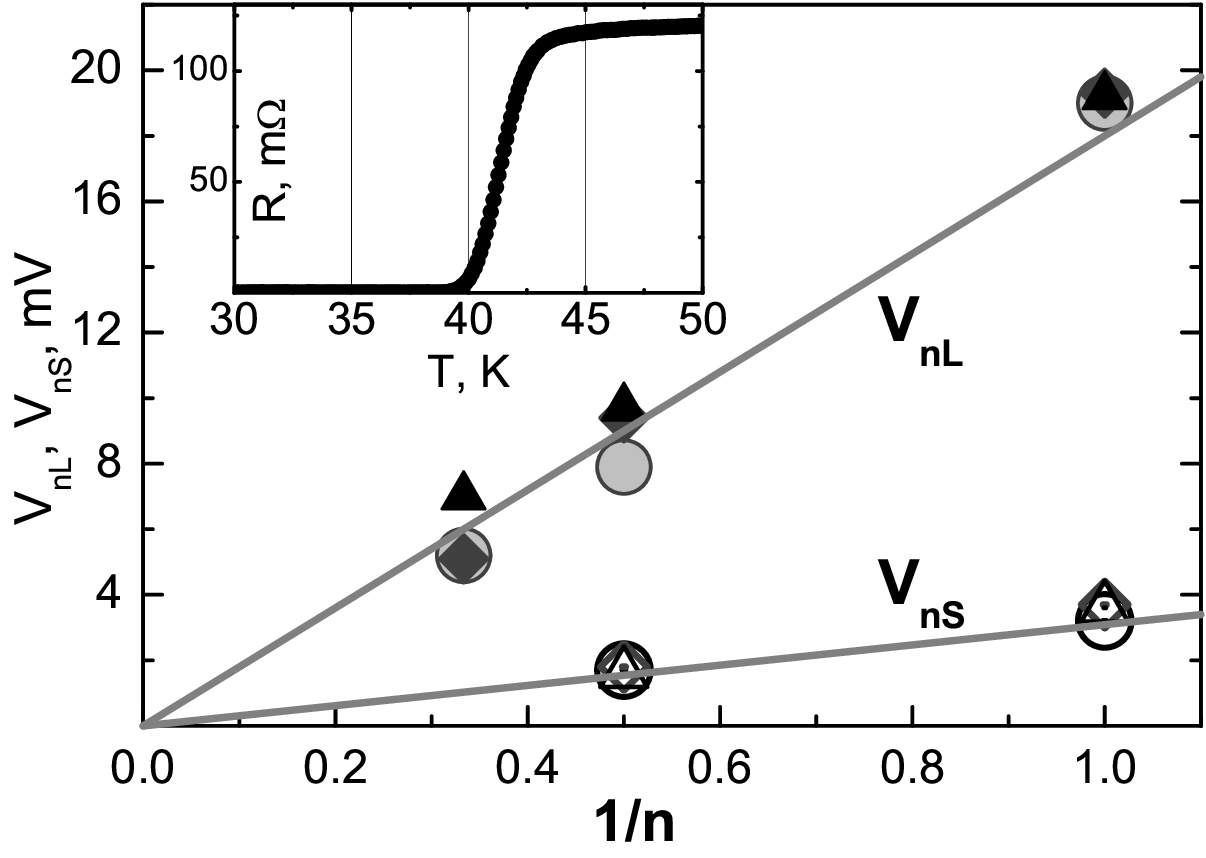}
\caption{
Minima positions $V_{nL} = 2\Delta_L/en_L$ and $V_{nS} = 2\Delta_S/en_S$ versus $1/n$ for the studied Andreev ScS nanocontacts (see Fig. 1): sample Ce1, contacts $\sharp d1$ (circles) and $\sharp d2$ (triangles), and sample Ce2, contact $\sharp d6$ (rhombs). The averaged gap values are $\Delta_L = 9 \pm 1$\,meV, $\Delta_S = 1.6 \pm 0.2$\,meV. \textbf{Inset} shows temperature dependence of the resistance through the superconducting transition for Ce-1111 sample ($T_C^{\rm bulk} = 41 \pm 1$\,K)
}
\label{Fig2}
\end{center}
\end{figure}

Dynamic conductance for three single ScS-contacts $\sharp d1$, $\sharp d2$ (marked as 1 and 2) for sample Ce1, and $\sharp d6$ (3) for sample Ce2 is shown in Fig. 1, where one can see  two sets of SGS. For comparison, the figure also shows the excess-current CVC for contact (3).  The dashed  areas comprise respective minima of the first set (marked with $n_L$ labels and originating from the large gap) and represent a 15\% uncertainty. Somewhat reduced intensity of the $n_L=1$ minima may be caused by a slight overheating of the contact area at the highest bias voltages. The fine structure in the bias voltage  interval between $n_L=1$ and $n_L=2$
signals a  large gap anisotropy of about 30\%.

The small gap SGS starts with minima located at $V_{S1} \approx \pm 3.3$\,mV (marked with dash-dotted vertical lines and $n_S$ labels) which have rather high relative amplitude, higher than the third Andreev dip from the large gap SGS. The sharp increase in the dip amplitude signals onset of a new SGS. Beyond the $n_S = 1$ dips one can also see the $n_S = 2$ dips located at $V_{S2} \approx \pm 1.6$\,mV.

The Andreev minima positions for the large and the small gap $V_{nL,S}$  are plotted in Fig. 2 as a function of $1/n$.  The plot clearly demonstrates the anticipated linear dependence which proves unambiguously that the dips in Fig. 1 do form two independent SGS, related with the presence of two superconducting gaps. The slope of the two fitting lines gives $\Delta_L = 9.0 \pm 1.4$\,meV, and $\Delta_S = 1.6 \pm 0.3$\,meV, for the large and small gaps, respectively. Taking into consideration the bulk $T_C = 41 \pm 1$\,K values (see inset to Fig.~2), we find the BCS ratio $2\Delta_L/k_BT_C^{\rm bulk} \approx 5.1$ for the  large gap, and $2\Delta_S/k_BT_C^{\rm bulk} \approx 0.9$ for the small gap.

\subsection{LaO$_{1-x}$F$_x$FeAs}
\label{La}

We studied about 50 ScS-Andreev contacts in polycrystalline LaO$_{0.9}$F$_{0.1}$FeAs (LOFA) samples with bulk $T_C^{\rm bulk}$ = $(28 \pm 1)$\,K. The dynamic conductance $dI(V)/dV$ of single contacts and nanosteps demonstrates two well-reproducible sets of SGS corresponding to the pair of independent gap values. The number $N$ of elementary contacts in  a stack was controlled by comparing the single contact $dI(V)/dV$ spectra with those for several stacks normalized to a single junction spectrum. Figure 3 shows the $dI(V)/dV$ spectra for a single contact $\sharp d17$ (black curve) and for the stacks $\sharp d9$ and $\sharp d10$ with various number of junctions in the array ($N = 2$, gray curve dI$_9$(V)/dV, and $N = 4$, dashed curve dI$_{10}$(V)/dV, respectively). Scaling of the SGS with properly selected number of contacts $N$ in nanosteps is straightforward. Following the equation of K\"{u}mmel {\it et al.} \cite{Kummel}, we easily obtain the large gap $\Delta_L \approx 6.1$\,meV. As for the small gap mi\-nima, peculiarities marked by arrows (at the top of Fig.~3 and in the inset) give $\Delta_S = 0.8$\,meV and 1.25\,meV values at $T = 4.2$\,K for the single contact $\sharp d17$ and for the array $\sharp d9$, respectively. By tracing the $\Delta_{L,S}(T)$ temperature dependence, we found the local critical temperature of the contact area $T_C^{\rm local}$. The  values
obtained,
$\Delta_L \approx 6.1$\,meV, $\Delta_S \approx 1.25$\,meV, and $T_C^{\rm local} = 26 \pm 1$\,K are close to the results of \cite{Yashima}.

\begin{figure}
\begin{center}
\includegraphics[width=0.46\textwidth]{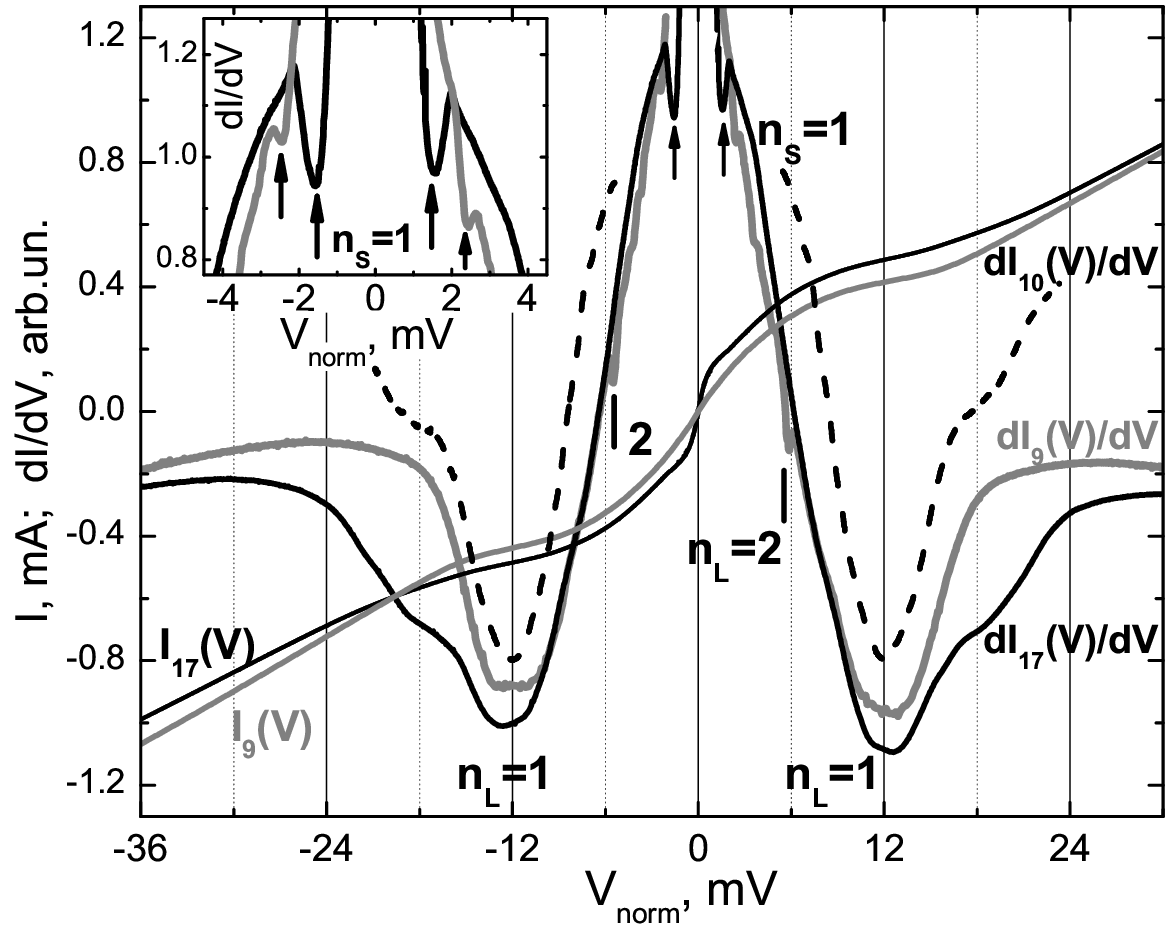}
\caption{La-1111. Normalized CVC and dynamic conductance of Andreev contacts for sample LOFA5: contacts $\sharp d10$ ($N = 4$ junctions in a stack, dashed line), $\sharp d9$ ($N = 2$, gray lines), both normalized to $\sharp d17$ (single junction, black lines). $T_C^{\rm local} \approx 26$\,K. The SGS for the large gap ($n_L$ labels) gives $\Delta_L \approx 6.1$\,meV for all the contacts, the small gap $\Delta_S \approx 0.8$\,meV (contact $\sharp d17$) and $1.25$\,meV (contact $\sharp d9$) SGSs are marked by black arrows and $n_S$ labels.
Inset blows-up details of the small gap SGS for  contacts $\sharp d17$ and $\sharp d9$
}
\label{Fig3}
\end{center}
\end{figure}

Figure 4 shows normalized CVC and dynamic conductance of two-contact ScS-Andreev array $\sharp d14$ in another LaO(F)FeAs sample.  The sharp SGS related to the large gap (marked with $n_L$ labels) gives $\Delta_L \approx 4.5$\,meV (see  solid squares in the inset to Fig. 4). Interestingly, all these Andreev minima up to $n_L = 5$ are double-splitted. It seems that the doublets observed on the high-quality characteristics are caused by some anisotropy of the $\Delta_L$ order parameter, though it has no nodes, as was shown in \cite{LOFA}. The CVC shown in Fig.~4 demonstrates a small Josephson supercurrent at zero bias caused by the tunneling between the sample clefts. An SGS associated with the small gap ($n_S$ labels and arrows) leads to $\Delta_S \approx 0.8$\,meV value (see  open squares in the inset to Fig. 4).

The BCS-ratio $2\Delta_L/k_BT_C^{\rm local} = (4.2 \div 5.6)$ exceeds the standard value 3.52 and thus is in favor of a strong coupling in  the $\Delta_L$ condensate. At the same time,   the small value, $2\Delta_S/k_BT_C^{\rm local} < 1.2$, is a result of induced superconductivity at finite temperatures in the bands with the small gap. These values support data reported earlier in \cite{LOFA} and are in close agreement with the experimental results on GdO(F)FeAs \cite{Gd} and our data on MgB$_2$ \cite{SSC,MgB2}.

\begin{figure}
\begin{center}
\includegraphics[width=0.44\textwidth]{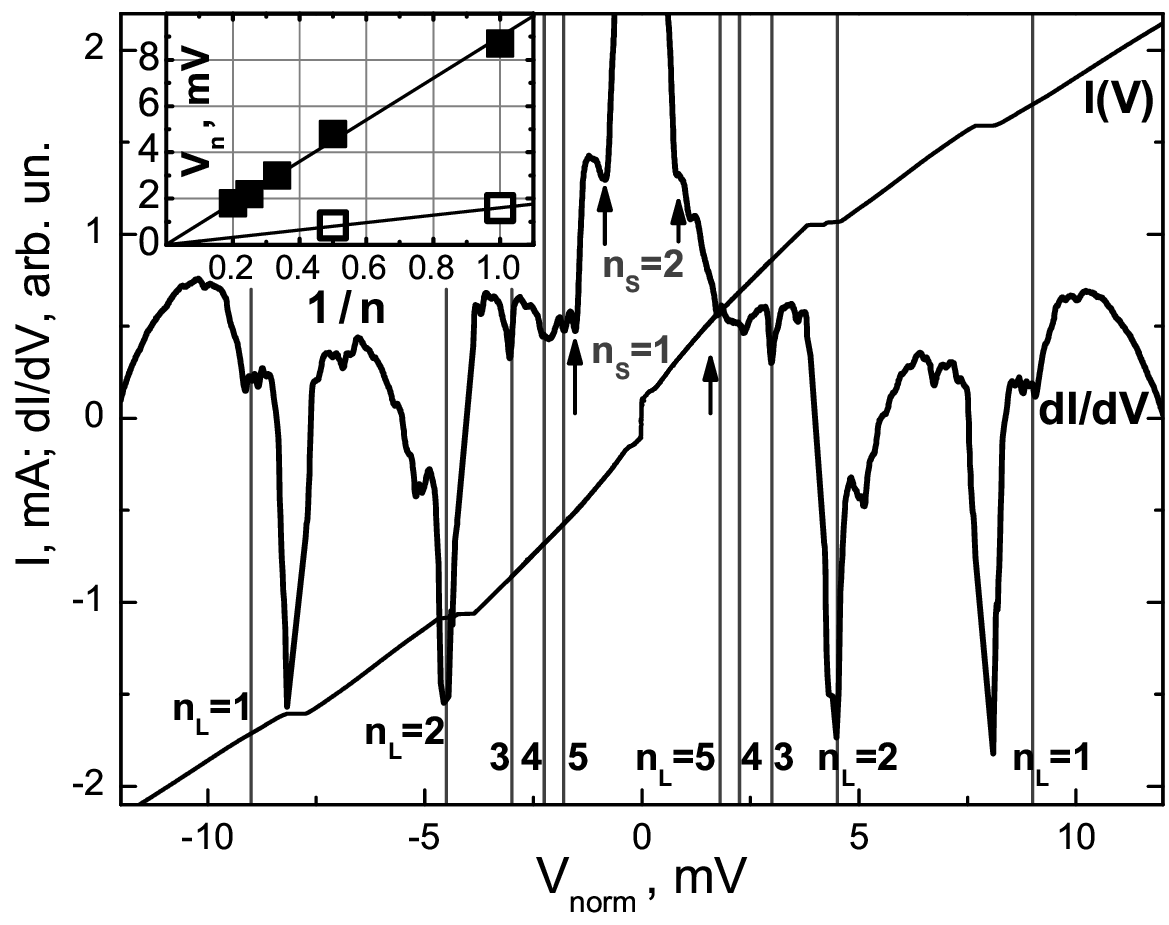}
\caption{La-1111. Normalized CVC and dynamic conductance  of  a two-contact Andreev array. The SGS of the large gap ($n_L$ labels) gives $\Delta_L \approx 4.5$\,meV ($T_C^{\rm local} \approx 25$\,K). Thin vertical lines represent the expected location of Andreev minima in accordance with theoretical formula $V_n = 2\Delta /en$ from \cite{Kummel}. The set of small gap peculiarities ($n_S$ labels and black arrows) lead to  $\Delta_S~\approx~0.8$\,meV. (\textbf{Inset}) The $V_n(1/n)$ dependences plotted for the SGS minima of both gaps (using the data from the main panel). Lines average the experimental values
}
\label{Fig4}
\end{center}
\end{figure}

\subsection{FeSe}
\label{FeSe}

Among the new class of Fe-based superconductors \cite{Kamihara}, layered FeSe has the simplest crystal structure and relatively low critical temperature $T_C$. Polycrystalline FeSe samples have been grown from melt by spontaneous nucleation. The synthesis process was described in detail in \cite{FeSe}. The intrinsic multiple Andreev reflections effect was observed in FeSe nanosteps earlier \cite{FeSe}.

\begin{figure}
\begin{center}
\includegraphics[width=0.46\textwidth]{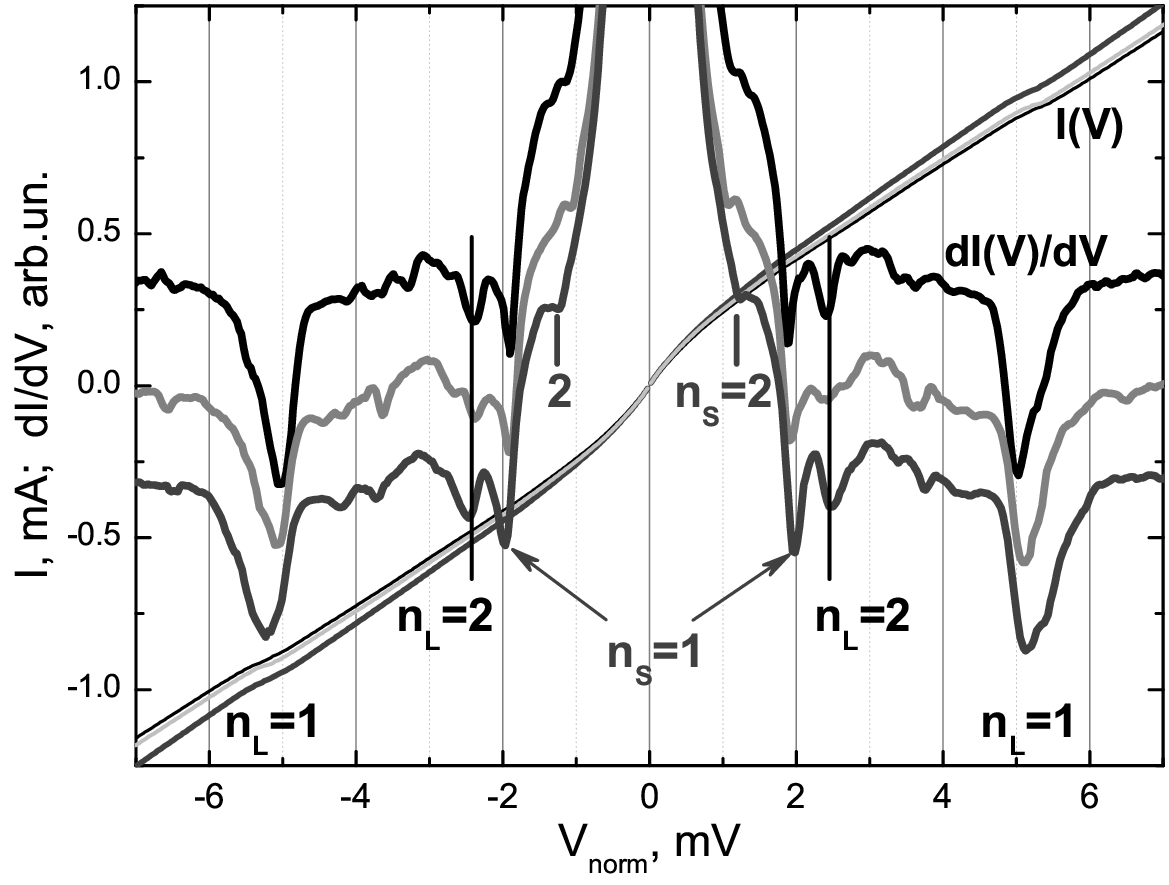}
\caption{FeSe. Normalized to a single junction CVC and $dI/dV$ spectra for ScS-Andreev contacts  (sample FS1, contacts $\sharp d8$, $\sharp d9$, $\sharp d11$ of three SnS-junctions in a stack; $T_C = 12.5$\,K, $T = 4.2$\,K). Two SGS at bias voltages $V_{nL,S}=2\Delta/en_{L,S}$ corresponding to the large ($n_L$ labels) and the small gap ($n_S$ labels) yield $\Delta_L \approx 2.6$\,meV and $\Delta_S~\approx~1$\,meV values.
}
\label{Fig5}
\end{center}
\end{figure}

The $I(V)$ and $dI(V)/dV$ characteristics for several ScS-junctions formed by successive mechanical readjustments of the contact are shown in Fig.~5. Two sets of SGS with a number of dips are clearly seen. The first set of dips ($n_L$ labels) gives the large gap value $\Delta_L \approx 2.6$\,meV. The second set of dips ($n_S$ labels) corresponds to the small gap $\Delta_S \approx 1$\,meV. It is worth noting that the dip positions and, consequently, the gap values remain unchanged under the readjustment of the contact. This proves the high homogeneity of the sample superconducting properties in the contact area. The superconducting gap values at $T = 4.2$\,K averaged over more than 30 ScS-contacts,  are $\Delta_L = 2.8 \pm 0.4$\,meV and $\Delta_S = 0.8 \pm 0.2$\,meV ($T_C^{\rm bulk} = 12 \pm 1$\,K). These results agree with the preliminary data obtained with similar samples \cite{FeSe}.

Figure 6 shows the $\Delta_{L,S}(T)$ temperature dependences for two ScS-contacts in FeSe. For the large gap, the $\Delta_L(T)$-curve lies slightly below the standard BCS-like dependence. For the small gap, the $\Delta_S(T)$ dependence deviates  essentially from the BCS-type curve and is in a good agreement with the calculations in \cite{Khasanov}. Knowing the local $T_C^{\rm local} \approx 9.7$\,K, one can calculate the BCS-ratio. For the large gap, we obtain $2\Delta_L/k_BT_C^{\rm local}$ $\approx 5.7$ which exceeds the  BCS value for a single-gap superconductor. On the other hand, for the small gap,  the $2\Delta_S/k_BT_C$ ratio is much smaller than 3.52. Such a behavior resembles the situation in MgB$_2$ \cite{MgB2,SSC12,Nicol} and, by parity of reasoning, can be explained by  the $k$-space proximity effect \cite{Golubov,Yanson} between two superconducting condensates, where the large gap condensate plays a ``driving'' role.

\begin{figure}
\begin{center}
\includegraphics[width=0.46\textwidth]{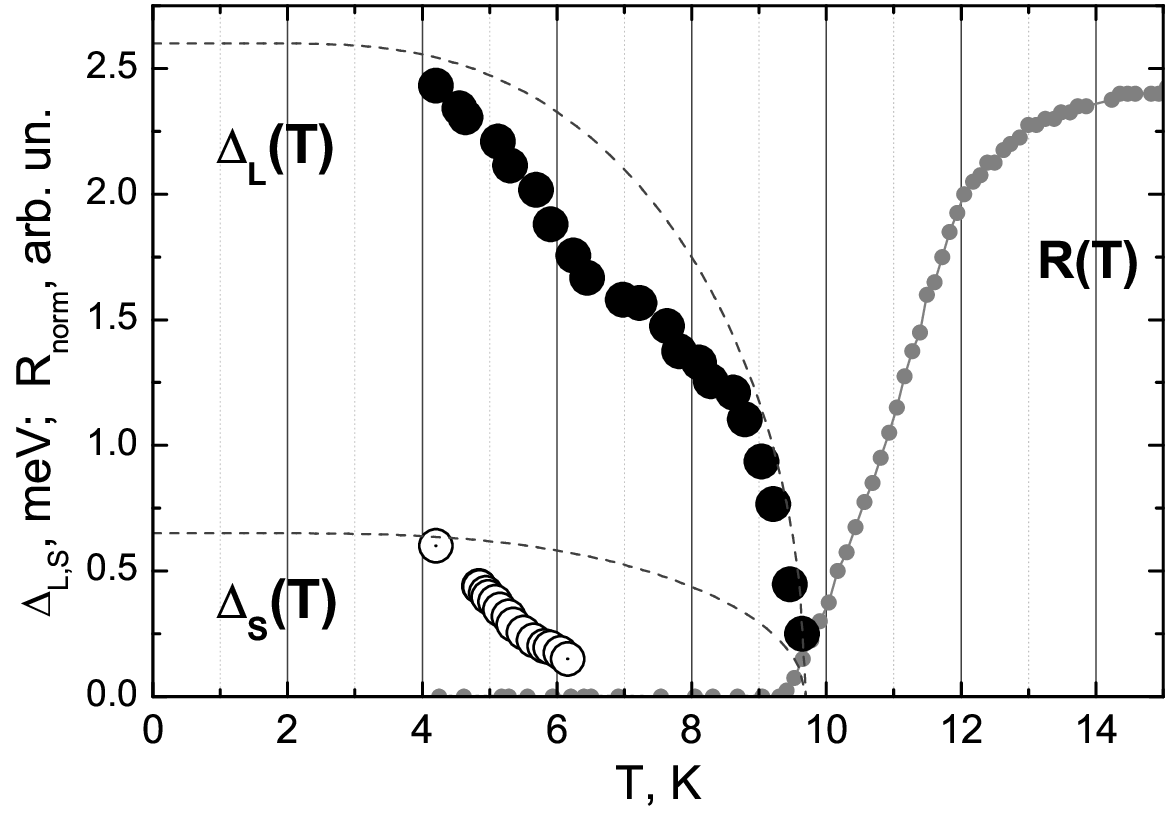}
\caption{Temperature dependence of two superconducting gaps  $\Delta_{L}(T)$ for FeSe (sample FS1, contact $\sharp$b; $\Delta_L(4.2~K) = 2.4 \pm 0.2$\,meV) and $\Delta_{S}(T)$ (sample FS2, contact $\sharp b$; $\Delta_S(4.2~K) = 0.7 \pm 0.1$\,meV). Local $T_C^{\rm local} = 9.7 \pm 0.5$\,K. Single-gap BCS-like curves (dashed lines), and the sample resistance $R_{\rm norm}(T)$ are shown for comparison
}
\label{Fig6}
\end{center}
\end{figure}

\section{Discussion}
\label{Discussion}

It was pointed out \cite{Onari} that inelastic neutron scattering data can provide a valuable information about the symmetry of the superconducting gap in novel superconductors. Calculations showed, in particular, that a hump structure must appear in the dynamic spin susceptibility just  above the $2\Delta$ energy in the case of an  $s^{++}$ wave state (the fully gapped $s$-wave state without sign reversal) \cite{Onari}. Recently, the experimental linear dependence of the spin resonance energy $E_{\rm res}$ on $T_C$ with the average slope $4.7k_BT_C$ was found for several iron based superconductors (see, e.g. Fig. 5 in \cite{Shamoto}). Within experimental errors, this dependence coincides with
 our
plot (Fig.~7) of the superconducting gap $2\Delta_L$ \emph{versus
 $T_C$ for several} iron based superconductors: Ce-1111 (present measurements), FeSe (\cite{FeSe}), LaO(F)FeAs \cite{LOFA}, GdO(F)FeAs \cite{Gd}, as well as KFe$_2$As$_2$, FeTe$_{1-x}$Se$_x$, LiFeAs, (see Fig. 11 in \cite{FeSe} and Refs. therein). Although the scattering of data in Fig. 7 is quite significant, two linear dependences emerge with $2\Delta_L/k_BT_C = 4.8 \pm 1.0$ and $2\Delta_S/k_BT_C = 1.1 \pm 0.4$. The coincidence of $2\Delta_L/k_BT_C$ (Fig. 7) and $E_{\rm res}/k_BT_C$ (Fig. 5 in \cite{Shamoto}) supports the version of a fully gapped $s$-wave state without sign reversal \cite{Onari}.

In conclusion, we studied properties of CeO(F)FeAs, LaO(F)FeAs, and FeSe superconductors by ScS-Andreev- and IMARE spectroscopies. The dynamic conductance curves for single and stack ScS-contacts cannot be described within the single-gap framework and evidence for the two-gap superconductivity in these compounds. For the first time studied CeO(F)FeAs ($T_C^{\rm bulk} \approx 41$\,K) we
 determined the two superconducting gaps $\Delta_L = 9.0 \pm 1.4$\,meV, and $\Delta_S = 1.6 \pm 0.3$\,meV; the respective BCS-ratios are $2\Delta_L/k_BT_C^{\rm bulk} \approx 5.1$, and $2\Delta_S/k_BT_C^{\rm bulk} \approx 0.9$.

For LaO(F)FeAs  ($T_C^{\rm bulk}\approx 28$\,K) we also determined the two superconducting  gap values $\Delta_L = 4.5 \div 6.5$\,meV, $\Delta_S = 0.8 \div 1.3$\,meV, leading to the BCS-ratios $2\Delta_L/k_BT_C^{\rm local} = 4.2 \div 5.6$ and $2\Delta_S/k_BT_C^{\rm local} = 0.6 \div 1.2$, respectively ($T_C^{\rm local} = 25 \div 29$\,K). We observed splitting of the SGS dips for high-quality characteristics, suggestive of an anisotropy of the $\Delta_L$ order parameter.

For FeSe ($T_C^{\rm bulk} \approx 12$\,K)  our IMARE spectroscopy data point to $\Delta_L = 2.8 \pm 0.4$\,meV,  $\Delta_S = 0.8 \pm 0.2$\,meV, and $2\Delta_L/k_BT_C^{\rm local} \approx 5.7$, $2\Delta_S/k_BT_C^{\rm local} \approx 1.5$. The temperature dependences $\Delta_{L,S}(T)$ indicate the $k$-space proximity effect between two superconducting condensates. The large gap BCS-ratio for all the materials studied exceeds 3.52, indicating a strong electron-boson coupling in the ``driving'' large gap condensate. The BCS-ratio for the small gap appears to be much less than 3.52, thus suggesting an induced superconductivity at finite temperatures in the ``driven'' $\Delta_S$ condensate due to a nonzero interband coupling. Finally, our data confirm a linear relation between the superconducting gap $\Delta_L$ and magnetic resonance energy $E_{\rm res} \approx 2\Delta_L$, valid for various Fe-based superconductors.

\begin{figure}
\begin{center}
\includegraphics[width=0.46\textwidth]{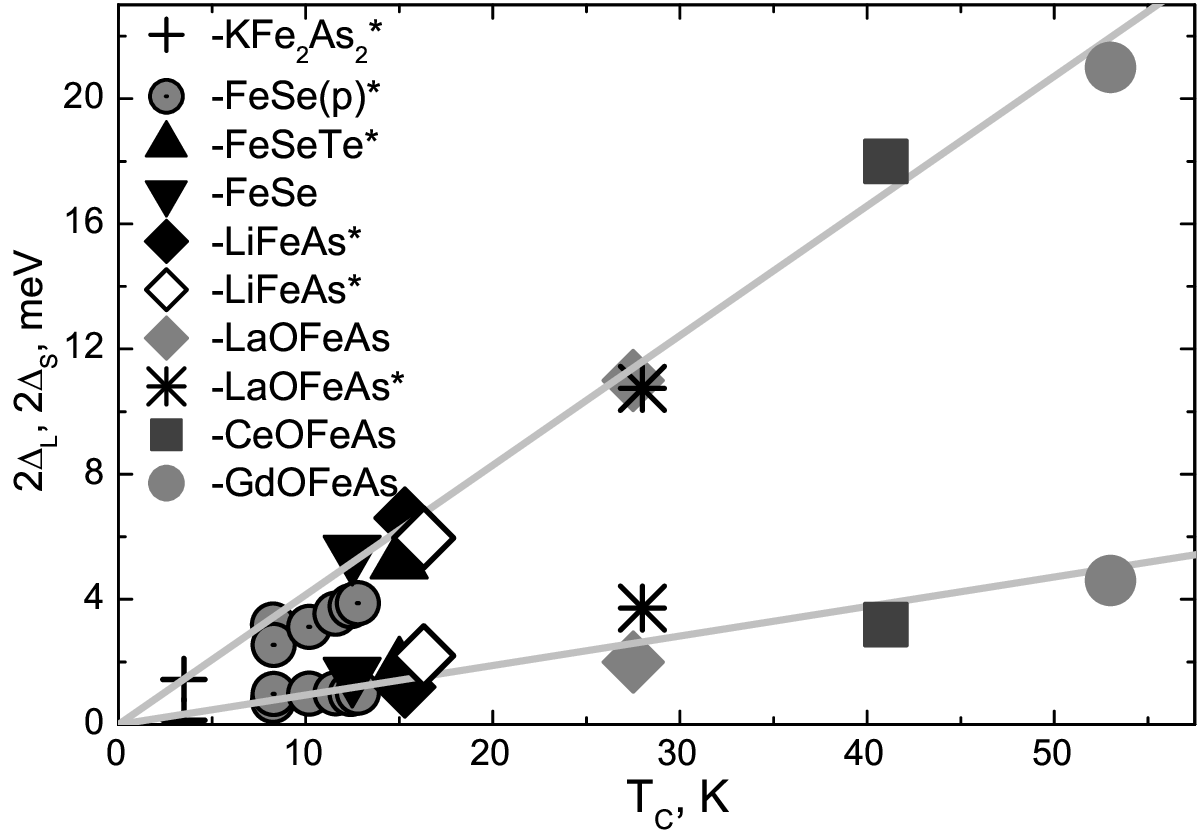}
\caption{
Scaling of the superconducting gap values with $T_C$ for iron-based superconductors: $2\Delta_L/k_BT_C = 4.8 \pm 1.0$ and $2\Delta_S/k_BT_C = 1.1 \pm 0.4$.
Asterisks
mark the data obtained by other groups (see Refs. in \cite{FeSe})
}
\label{Fig7}
\end{center}
\end{figure}

\begin{acknowledgements}
The work was supported by Russian Ministry of Education and Sciences (contract 11.519.11.60.12,
grant 8375), RFBR (grants 12-02-31269, 13-02-01451), DFG Grants 436RUS113 and FOR 538/BU887/4, and DFG priority program (SPP1458). We thank T. H\"{a}nke, C. Hess, G. Behr, R. Klingeler and B. B\"{u}chner for the La-1111 samples synthesis.
\end{acknowledgements}

\end{document}